\documentclass[aps,pra,twocolumn,superscriptaddress]{revtex4-1}

\usepackage{graphicx,color}
\usepackage{amssymb}
\usepackage{amsmath}

\begin{document}

\title{Plane wave coupling formalism for T-matrix simulations of light scattering by non-spherical particles}

\author{Dominik Theobald}
\email[]{dominik.theobald@kit.edu}
\affiliation{Light Technology Institute, Karlsruhe Institute of Technology, 76131 Karlsruhe, Germany}
\author{Amos Egel}
\affiliation{Light Technology Institute, Karlsruhe Institute of Technology, 76131 Karlsruhe, Germany}
\affiliation{Institute of Microstructure Technology, Karlsruhe Institute of Technology, 76131 Karlsruhe, Germany}
\author{Guillaume Gomard}
\affiliation{Light Technology Institute, Karlsruhe Institute of Technology, 76131 Karlsruhe, Germany}
\affiliation{Institute of Microstructure Technology, Karlsruhe Institute of Technology, 76131 Karlsruhe, Germany}
\author{Uli Lemmer}
\affiliation{Light Technology Institute, Karlsruhe Institute of Technology, 76131 Karlsruhe, Germany}
\affiliation{Institute of Microstructure Technology, Karlsruhe Institute of Technology, 76131 Karlsruhe, Germany}

\date{\today}

\begin{abstract}
The computation of light scattering by the superposition T-matrix scheme has been so far restricted to systems made of particles that are either sparsely distributed or of near-spherical shape. In this work, we extend the range of applicability of the T-matrix method by accounting for the coupling of scattered fields between highly non-spherical particles in close vicinity. This is achieved using an alternative formulation of the translation operator for spherical vector wave functions, based on a plane wave expansion of the particle's scattered electromagnetic field. The accuracy and versatility of the present approach is demonstrated by simulating arbitrarily oriented and densely packed spheroids, for both, dielectric and metallic particles.
\end{abstract}

\maketitle

\section{Introduction}
The quantitative description of light scattering by wavelength-scale particle systems is of paramount importance for a wealth of disciplines. Examples span from the characterization and sensing of atmospheric particulates \cite{geier2014}, the performance of astrophysical studies \cite{Tamanai2006}, investigations in biology \cite{Wilts2017} and biomedicine \cite{Dannhauser2017}, to the optimization of light scattering for plasmonic devices \cite{Atwater2010}, light-emitting diodes \cite{Gomard2016} and solar cells \cite{Zhang2011}.

Strictly numerical simulation techniques, as the finite element method (FEM) and the finite difference time domain (FDTD) method, provide comfortable solutions for small or periodic systems. However, for larger disordered photonic systems, they require enormous computational resources, rendering the treatment of larger complex systems impossible. An efficient alternative to these numerical tools can be provided by the T-matrix method \cite{Waterman1965,Doicu2006,Mishchenko2002} in conjunction with the translation addition theorem of the spherical vector wave functions \cite{Cruzan1962} to account for multiple scattering. However, the applicability of this approach is so far limited to ensembles that are either sparsely distributed, or that consist of particles with nearly spherical shape, whereas it breaks down for systems of non-spherical particles, when the particle inter-distance is low. In fact, the well-established superposition T-matrix scheme \cite{Peterson1973,Varadan1980,Mishchenko1996} for multi-particle systems requires that any particle's circumscribing sphere does not intersect adjacent particles. One attempt to overcome this limitation is to decompose a single scatterer into multiple sub-units, which are then treated as a multiple scattering problem \cite{Wriedt2008a}. This way, the downsized sub-units' circumscribing spheres exhibit less overlap.

In this contribution, we develop an alternative formalism to accurately describe the multiple scattering between close-by non-spherical particles. The basic idea is to transform the scattered field's spherical wave expansion (SWE) into a plane wave expansion (PWE), allowing the use of the much simpler plane wave translation addition theorem instead of the spherical wave translation addition theorem. This way, the non-overlap restriction of the particles' circumscribing spheres can be avoided, provided that for each pair of particles a separating plane can be found. This is always the case for convex particles. The concept extends our recent work for the case of an oblate particle near a planar interface \cite{Egel2016b}.

We briefly summarize in Sec. \ref{sec:general scheme} the procedure of the superposition T-matrix scheme and highlight its range of applicability. In Sec. \ref{sec:planewavecoupling}, we give a comprehensive description of the plane wave coupling formalism for arbitrary orientation of the involved scattering particles.

Finally in Sec. \ref{sec:examples}, we study light scattering at two exemplary configurations, both for dielectric and metallic spheroids. We compare results computed with both, the conventional procedure based on the SWE translation addition theorem and the new plane wave coupling formalism to reference simulations using the FEM. The first example is given by a two-spheroid configuration, and the second example shows a dense cluster of twenty nano-rods, which are utilized, e.g., for light management in photovoltaics \cite{Shital2016}.

\section{Scattering by multiple particles: general T-matrix formalism \label{sec:general scheme}}
One of the most powerful tools to study light scattering by nano-particles is the T-matrix method \cite{Waterman1965}. For clarity, we briefly summarize its procedure. Comprehensive descriptions can be found, e.g., in Ref(s). \cite{Doicu2006,Mishchenko2002}.

We consider a single particle in a homogeneous, isotropic, linear and nonabsorbing  medium. The electric field $\mathbf{E}(\mathbf{r})$ can be expressed as a superposition of an incoming electric field $\mathbf{E}_{\mathrm{in}}(\mathbf{r})$ and a scattered electric field $\mathbf{E}_{\mathrm{sc}}(\mathbf{r})$:
\begin{equation}
\mathbf{E}(\mathbf{r})=\mathbf{E}_{\mathrm{in}}(\mathbf{r})+\mathbf{E}_{\mathrm{sc}}(\mathbf{r}) .
\end{equation}
In the T-matrix formalism, the incoming electric field is written as a sum of regular spherical vector wave functions (SVWFs) $\mathbf{M}_n^{(1)}(\mathbf{r})$:
\begin{equation}
\mathbf{E}_{\mathrm{in}}(\mathbf{r}) =\displaystyle\sum_{n} a_n \mathbf{M}_n^{(1)}(\mathbf{r}) .
\end{equation}
The scattered field is written in outgoing SVWFs $\mathbf{M}_n^{(3)}(\mathbf{r})$:
\begin{equation}
\mathbf{E}_{\mathrm{sc}}(\mathbf{r}) =\displaystyle\sum_{n} b_n \mathbf{M}_n^{(3)}(\mathbf{r}) .
\end{equation}
Here, the summation index $n$ subsumes the degree $l$ and order $m$ of the multipole, as well as the polarization $p$ of the spherical wave, $n=(l,m,p)$.

The T-matrix of a scattering particle $S$ is defined as the linear operator that maps the amplitudes of incoming wave functions $a_n^S$ to the amplitudes of outgoing wave functions $b_n^S$:
\begin{equation}
\label{equ:tmat}
b_n^S = \displaystyle\sum_{n'} T_{nn'} a_{n'}^S.
\end{equation}
$T$ contains the complete information about the scattering properties of a particle. Accurate and time efficient methods for the computation of the T-matrix are available for a broad variety of scattering particles (for a  collection of computer codes, see for example the information portal described in Ref. \cite{Wriedt2008}). In the following, we assume that the T-matrix of each particle is precisely known.

Increasing the particle number to at least two results in a system where multiple scattering has to be taken into account. Such systems have been well studied in the framework of the superposition T-matrix scheme \cite{Liu2008,Mackowski2011,Egel2014,Fruhnert2016,Markkanen2017}.
Multiple scattering denotes, that each particle's scattered field $\mathbf{E}_{sc}^{S'}(\mathbf{r})$ contributes to the incoming field $\mathbf{E}_{in}^{S}(\mathbf{r})$ of particle $S$. Since the incoming field at a particle $S$ is not known anymore, (\ref{equ:tmat}) is not sufficient to describe the entire system. A second equation is needed to determine the incoming field for each particle:
\begin{equation}
a_n^S = a_n^{S,\mathrm{in}} + \displaystyle\sum_{S\neq S'} \displaystyle\sum_{n'} W_{nn'}^{SS'} b_{n'}^{S'} .
\label{equ:wmat}
\end{equation}
Here, $a_n^{S,\mathrm{in}}$ denotes the amplitudes of the initial incoming spherical waves at particle $S$, which are generated by the initial field excitation, e.g., a plane wave or a dipole field. The coupling matrix $\mathrm{W}^{SS'}$ describes, how the scattered field of particle $S'$ contributes to the incoming field of particle $S$. It corresponds to the transpose of the translation operator $A$:
\begin{equation}
W_{nn'}^{SS'} = A_{n'n}(\mathbf{r}_S-\mathbf{r}_{S'}) .
\end{equation}
For the translation addition theorem of SVWFs (see (\ref{equ:translationadditiontheorem})),
the translation operator can be expressed either in a closed form expression involving the Wigner-3j symbols \cite{Mishchenko2002} or be constructed from an iterative scheme, which can be found in Ref. \cite{Doicu2006}.

\begin{figure}
	\def\svgwidth{8.6cm}
	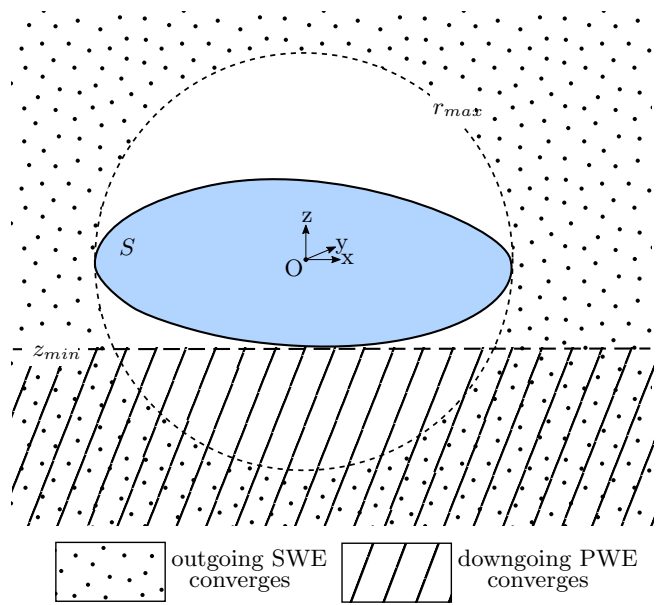	
	\caption{\label{fig:domainsofvalidity}Oblate scattering particle $S$ in a homogeneous medium. The SWE of the particle's scattered field $\mathbf{E}_{sc}^S(\mathbf{r})$ is only valid outside the particle's smallest circumscribing sphere ($r > r_{max}$). A correct PWE can be obtained everywhere below the particle ($z < z_{min}$).}
\end{figure}

Inserting (\ref{equ:wmat}) into (\ref{equ:tmat}) results in a self-consistent set of equations to account for light scattering by a system of multiple particles. Written in a matrix-vector notation we obtain:
\begin{equation}
\mathbf{b} = (1-\mathbf{TW})^{-1}\mathbf{Ta}^{\mathrm{in}} .
\label{equ:mainequ}
\end{equation}
Any particle system built by spheres can be described by Eq. (\ref{equ:mainequ}). However, for different particle shapes we have to restrict ourselves to configurations where the distance between particles is large enough to ensure that a particle's circumscribing sphere does not overlap with any other particle. The scattered field's SWE is only valid outside the particle's smallest circumscribing sphere (see Fig. \ref{fig:domainsofvalidity}). Inside the circumscribing sphere, the field expansion may not converge towards its true value \cite{Bates1975}. This restriction can be slightly relaxed towards a sphere, circumscribing the singularities of the scattered field expansion \cite{Doicu1999}. The question arises, how a correct field representation in the dashed white region can be achieved. 

\section{Near field coupling of non-spherical particles via plane waves \label{sec:planewavecoupling}}
Utilizing an example of light scattering at a particle near a finite cylinder, Bostr\"om et al. \cite{Bostrom1991} suggested the idea of transforming between spherical, cylindrical and plane wave representations, whenever one of them is not suitable for the configuration considered. Following this idea, we propose to make use of a plane wave representation of the scattered fields to overcome the separation restriction of the superposition T-matrix formalism. 

The benefit of transforming the outgoing SWE into a PWE has been recently shown for non-spherical particles close to a layer interface \cite{Egel2016b}. In short, the intermediate transformation of the SWE into a truncated PWE acts as a regularization of the divergent part of the SWE in the near field region. In fact, the domain of validity for a down going PWE is limited by a plane that is tangential to the particle from below and oriented such that its normal coincides with the $z$-direction (see dashed region in Fig. \ref{fig:domainsofvalidity}), thereby allowing a correct representation of the scattered field nearby the particle, where the SWE would diverge. This holds, even if the PWE is constructed starting from a divergent SWE.

\begin{figure}
	\def\svgwidth{8.6cm}
	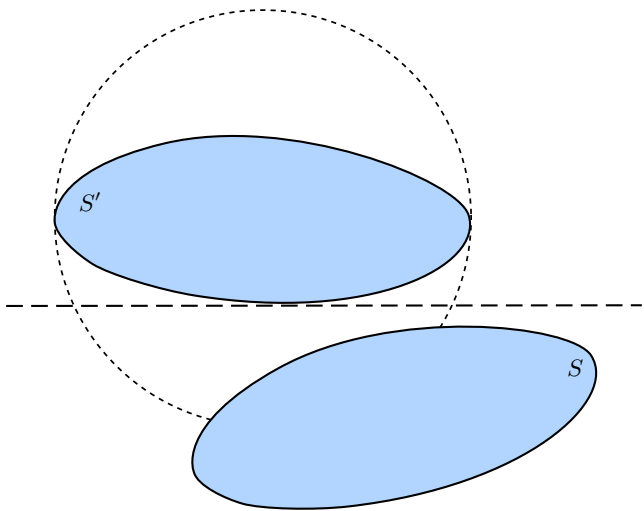
	\caption{\label{fig:2spheroidsclosedistance}Two oblate scattering particles at a close distance: particle $S$ intersects the circumscribing sphere of particle $S'$, but it is entirely below the bounding plane.}
\end{figure}

For simplicity, we consider a system of two non-spherical particles at a close distance (see Fig. (\ref{fig:2spheroidsclosedistance})). Each particle intersects with the other's circumscribing sphere. Therefore, the conventional T-matrix formalism based on the translation addition theorem is in general not suitable to model such configurations.
Note that for the depicted configuration, the lower particle does not intersect the upper particle's bounding plane, such that it is entirely located in the domain where the PWE of the scattered field from the upper particle is valid. 

To circumvent the limitation of the conventional approach based on the translation addition theorem, we thus introduce a formalism to couple $\mathbf{E}_{\mathrm{sc}}^{S'}$ to $\mathbf{E}_{\mathrm{in}}^{S}$ in terms of a PWE, including three main steps: 
\begin{itemize}
	\item a transformation of the (outgoing) SWE of $\mathbf{E}_{\mathrm{sc}}^{S'}$ into a PWE,
	\item a translation of the PWE of $\mathbf{E}_{\mathrm{sc}}^{S'}$ to the center of particle $S$,
	\item a retransformation of the PWE of $\mathbf{E}_{\mathrm{sc}}^{S'}$ into a (regular) SWE of $\mathbf{E}_\mathrm{in}^{S}$.
\end{itemize}
As pointed out in the previous section, the incoming field of a single particle is not explicitly known in a multi-particle system, preventing direct field transformations. Instead, we aim for a PWE formulation of the translation operator $A$. 

We start with an outgoing SVWF $\mathbf{M}_n^{(3)}(\mathbf{r}-\mathbf{r}_{S'})$ with its center at position $\mathbf{r}_{S'}$. Expanding it in terms of a down going plane wave (\ref{equ:EtoM3}) and translating it to a position $\mathbf{r}_{S}$ results in:
\begin{align}
\mathbf{M}_n^{(3)}(\mathbf{r}-\mathbf{r}_{S'}) = \ & \frac{1}{2\pi} \displaystyle\int_{\mathbb{R}^2}d^2\mathbf{k}_\parallel \frac{1}{k_zk} \displaystyle\sum_{j=1}^{2} B_{nj}\left(\frac{-k_z}{k}\right) \notag \\
& \times \mathrm{e}^{\mathrm{i}m\alpha} \mathrm{e}^{\mathbf{k}\cdot(\mathbf{r}_{S}-\mathbf{r}_{S'})} \mathbf{E}_j^-(\kappa,\alpha;\mathbf{r}-\mathbf{r}_{S}) . 
\end{align}
In our notation $\mathbf{E}^-$ refers to a plane wave propagating in negative z-direction (see (\ref{equ:planewave})) and $B$ denotes the transformation operator between spherical waves and plane waves (\ref{equ:Boperator}). 

Utilizing (\ref{equ:M1toE}), we retransform the plane wave into a regular spherical wave:
\begin{align}
\label{equ:M3inM1}
\mathbf{M}_n^{(3)}(\mathbf{r}-\mathbf{r}_{S'}) = \ & \frac{2}{\pi} \displaystyle\int_{\mathbb{R}^2}d^2\mathbf{k}_\parallel \frac{1}{k_zk} \notag \\
& \times \displaystyle\sum_{j=1}^{2} B_{nj}\left(\frac{-k_z}{k}\right) \mathrm{e}^{\mathrm{i}m\alpha} \mathrm{e}^{\mathbf{k}\cdot(\mathbf{r}_{S}-\mathbf{r}_{S'})} \notag \\
& \times \displaystyle\sum_{n'}  B_{n'j}^\dagger\left(\frac{-k_z}{k}\right) \mathrm{e}^{-\mathrm{i}m'\alpha} \mathbf{M}_{n'}^{(1)}(\mathbf{r}-\mathbf{r}_{S}) . 
\end{align}
By comparing (\ref{equ:M3inM1}) with the translation addition theorem for SVWFs (\ref{equ:translationadditiontheorem}), we obtain a formulation of the translation operator $A_{n'n}(\mathbf{r}_{S}-\mathbf{r}_{S'})$, based on a plane wave expansion.

Writing out $\int d^2\mathbf{k}_\parallel=\int d\kappa\kappa \int d\alpha$ and utilizing $\mathbf{k}\cdot(\mathbf{r}_{S}-\mathbf{r}_{S'}) = \kappa\rho_{SS'}\cos(\alpha-\varphi_{SS'})+k_zz_{SS'}$ with $(\rho_{SS'},\varphi_{SS'},z_{SS'})$ being the cylindrical coordinates of $(\mathbf{r}_{S}-\mathbf{r}_{S'})$, we obtain:
\begin{align}
\label{equ:doubleintformulation}
W_{nn'}^{SS'} &= A_{n'n}(\mathbf{r}_{S}-\mathbf{r}_{S'}) \notag \\
&= \frac{2}{\pi} \displaystyle\sum_{j=1}^{2} \displaystyle\int d\kappa \frac{\kappa}{k_zk}  B_{nj}\left(\frac{-k_z}{k}\right) B_{n'j}^\dagger\left(\frac{-k_z}{k}\right)  \notag \\
&\ \ \ \times \mathrm{e}^{\mathrm{i}(-k_zz_{SS'})} \displaystyle\int d\alpha \: \mathrm{e}^{\mathrm{i}(\kappa \rho_{SS'} \cos(\alpha-\varphi_{SS'}))} \mathrm{e}^{\mathrm{i}\alpha(m-m')} . 
\end{align}
To get rid of the double integral, one can compare the second integral in (\ref{equ:doubleintformulation}) with the integral formulation of the Bessel function $J$ reported in Ref. \cite{Adams}:
\begin{equation}
J_a(\varrho) = \frac{\mathrm{i}^{-a}}{2\pi} \displaystyle\int_{-\pi}^{\pi} \mathrm{e}^{\mathrm{i}\varrho\cos\phi+\mathrm{i}a\phi} d\phi .
\end{equation}
Finally, we end up with:
\begin{align}
\label{equ:Wfinal}
W_{nn'}^{SS'} = \ & 4\mathrm{i}^{m-m'} \displaystyle\sum_{j=1}^{2} \displaystyle\int d\kappa \frac{\kappa}{k_zk} B_{nj}\left(\frac{-k_z}{k}\right) B_{n'j}^\dagger\left(\frac{-k_z}{k}\right) \notag \\
& \times \mathrm{e}^{\mathrm{i}(-k_zz_{SS'})} \mathrm{e}^{\mathrm{i}\varphi_{SS'}(m-m')} J_{m-m'}(\kappa\rho_{SS'}) . 
\end{align}

Note that a transformation-translation-transformation scheme to utilize the simple plane wave addition theorem for SVWF translations has previously been suggested in Ref. \cite{Bostrom1991}. However, to the best of our knowledge, the use of this method in order to regularize divergent near-field SWE in the context of multiple scattering by means of a truncation of the involved PVWF wavenumbers has not been reported before.

So far, we have introduced a formalism to couple the scattered electric field of one particle to another by transforming the outgoing SWE into a PWE at a plane, parallel to the xy-plane ($z=0$) of our laboratory coordinate system ($L$). In a more general case, a plane separating two adjacent particles will not be parallel to the xy-plane, but arbitrarily aligned in space. Then, one can perform the plane wave coupling formalism in a rotated coordinate system ($R$), in which the separation plane is parallel to the xy-plane.

\begin{figure}
	\def\svgwidth{8.6cm}
	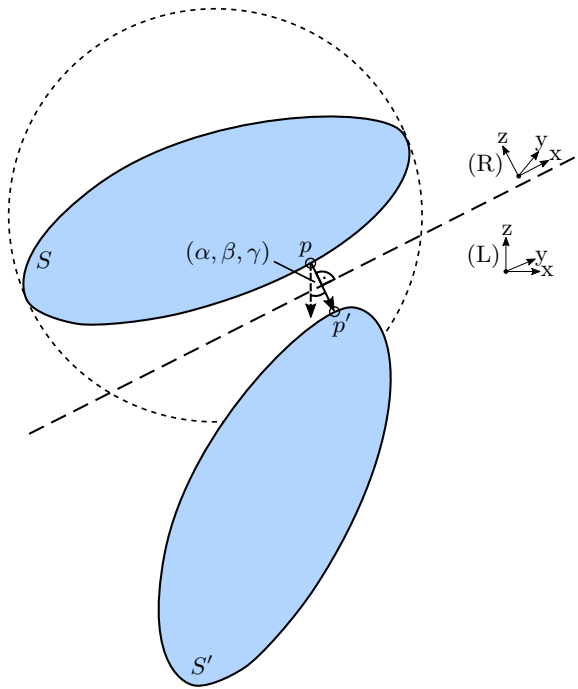
	\caption{\label{fig:seperationplane}A plane separating the particles $S$ and $S'$ is normal to the vector $\overline{pp'}$, connecting the two surface points $p$ and $p'$ that are closest to each other. A rotation of $\overline{pp'}$ towards the z-vector $\hat{e}_z$ of the laboratory coordinate system $(L)$ by the Euler angles $(\alpha,\beta,\gamma)$ ensures that the separation plane is parallel to the xy-plane of the rotated coordinate system $(R)$.}
\end{figure}

Let $\mathbf{D}$ be a matrix notation of the rotation addition theorem (Eq. (\ref{equ:rotationadditiontheorem})) with
\begin{equation}
D_{lmpl'm'p'}(\alpha,\beta,\gamma) = D_{mm'}^l(\alpha,\beta,\gamma)\delta_{ll'} . \notag
\end{equation}
Then we obtain the coupling matrix in the laboratory coordinate system of particles $S$ and $S'$ in terms of the coupling matrix in the rotated coordinate system:
\begin{equation}
\mathbf{W}^{SS'}(L) = \mathbf{D}^T(-\gamma,-\beta,-\alpha) \mathbf{W}^{SS'}(R) \mathbf{D}^T(\alpha,\beta,\gamma) .
\end{equation}
To determine the angles of rotation, one needs to find a plane separating the two particles. It is assured that such a plane exists if two particles, with a convex surface shape, do not touch or overlap. One way to obtain such a plane is to find the two surface points $p$ and $p'$ on particle $S$ and $S'$ that are closest to each other. Then, the separation plane is simply perpendicular to the vector $\overline{pp'}$, as illustrated in Fig. \ref{fig:seperationplane}. Since we want the separation plane to be parallel to the xy-plane in our rotated coordinate system, the angles $(\alpha,\beta,\gamma)$ have to transform $\overline{pp'}$ into $|\overline{pp'}|\hat{e}_z$ in the laboratory coordinate system.

To conclude, we have introduced a formalism that couples the scattered electric field of a particle to another by making use of a plane wave representation. An accurate near field representation of the scattered field can thereby be achieved in a region, where the SWE of the scattered field is not valid. Performing the plane wave coupling in a rotated coordinate system allows accounting for light scattering by any pair of arbitrarily oriented, non-spherical particles. 

\section{Application examples \label{sec:examples}}
In this section, we evaluate the accuracy of T-matrix simulations relying on the plane wave coupling formalism, as  introduced in Sec. \ref{sec:planewavecoupling}. To this end, we compare them with simulations performed with the well-established FEM (available in the Comsol Multiphysics$^{\circledR}$ software), used as a benchmark. The suitability of the present approach is also emphasized by introducing results obtained with the conventional superposition T-matrix scheme (see Sec. \ref{sec:general scheme}). To demonstrate the generality of the plane wave coupling formalism, we consider both (lossless) dielectric and (lossy) metallic nano-particles, and scattering systems either based on two particles or on a cluster made of twenty particles.  In the following, all configurations are excited by a plane wave $(\lambda=500 \ \mathrm{nm})$, which is polarized along the y-direction and propagating in negative z-direction. The ambient medium is chosen to be air, $n_a = 1$.

\subsection{Two arbitrarily oriented particles}
\begin{figure*}
	\def\svgwidth{17.8cm}
	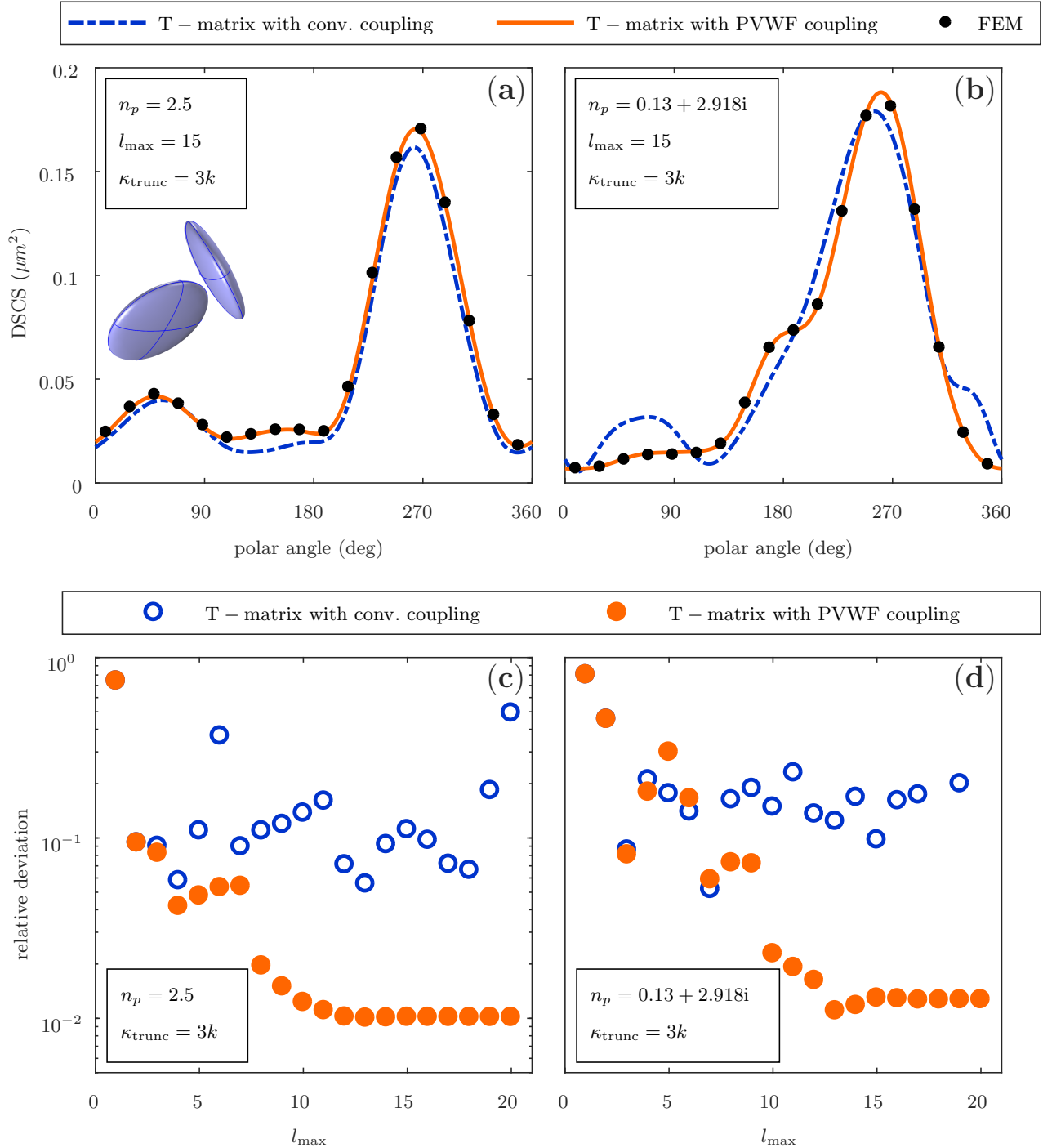
	\caption{\label{fig:2spheroidsplotgroup}Light scattering at two oblate spheroids with semi-axes $(a=b=200 \ \mathrm{nm}, c=50 \ \mathrm{nm})$. The particles are excited by a plane wave $(\lambda=500 \ \mathrm{nm})$, polarized in y-direction and propagating in negative z-direction. The ambient medium is air $(n_a=1)$. (a) DSCS of two $\mathrm{TiO_2}$-particles $(n_p=2.5)$. For the SWE, multipole orders up to $l_{max}=15$ are considered. The PWE is truncated at $\kappa_{\mathrm{trunc}}=3k$. (b) DSCS of two oblate $\mathrm{Ag}$-particles $(n_p=0.13+2.918\mathrm{i})$. The relative deviation of the DSCS for the conventional T-matrix formalism and the PVWF coupling procedure with respect to the FEM simulations is shown for $\mathrm{TiO_2}$ (c) and $\mathrm{Ag}$ (d). The maximal multipole order is varied from $l_{max}=1$ up to $l_{max}=20$, while the PWE truncation is kept constant at $\kappa_{\mathrm{trunc}}=3k$.}
\end{figure*}

In a first step, we study light scattering by a system consisting of two particles, which are either made of a dielectric $(\mathrm{TiO_2})$ or of a metallic $(\mathrm{Ag})$ material. The scattering particles considered are oblate spheroids, with semi-major axes of $a=b=200 \ \mathrm{nm}$ and a semi-minor axis of $c=50 \ \mathrm{nm}$, corresponding to dimensionless size parameters of $ka=kb=2.51$ and $kc=0.63$. The refractive index of $\mathrm{TiO_2}$ $(n_p=2.5)$ corresponds to the bulk value of titania in the anatase phase and at a vacuum wavelength of $\lambda=500 \ \mathrm{nm}$ \cite{Jalava2015}. We note that the refractive index of nano-particles can strongly deviate from its material's bulk value. However, the here used relatively large particle diameter justifies the use of $\mathrm{TiO_2}$'s bulk property. The first particle's center is placed at $(x_1= -80 \ \mathrm{nm}, y_1= 25 \ \mathrm{nm}, z_1= 120 \ \mathrm{nm})$, while the second particle's center is located at $(x_2 = 120 \ \mathrm{nm}, y_2 = -20 \ \mathrm{nm}, z_2 =-60 \ \mathrm{nm})$. The particles' orientations are obtained by rotation of $(\alpha_1=\frac{8}{9}\pi, \beta_1=\frac{1}{3}\pi)$ and $(\alpha_2=\frac{14}{9}\pi, \beta_2=\frac{5}{18}\pi)$ with respect to a spheroid with its semi-minor axis directed along the z-axis. In this case, the minimal inter-particle distance measures $18 \ \mathrm{nm}$. A visualization of the configuration can be found as an inset in Fig. \ref{fig:2spheroidsplotgroup}(a).

Figure \ref{fig:2spheroidsplotgroup} shows the comparison between the conventional superposition T-matrix scheme, for which the translation operator $A$ (see (\ref{equ:translationadditiontheorem})) has been computed by making use of the Wigner-3j symbols, and the T-matrix scheme relying on the plane wave coupling formalism (Sec. \ref{sec:planewavecoupling}). For the computation of all T-matrices a Fortran code based on the null-field method with discrete sources (NFM-DS)  \cite{Doicu2006} has been used. For reference, we compare our results to FEM simulations. In Fig. \ref{fig:2spheroidsplotgroup}(a), the differential scattering cross-section (DSCS) in the yz-plane is shown. In this example, the SWE has been performed up to a maximal multipole order of $l_{\mathrm{max}}=15$. A substantial deviation of the blue-dotted line from the FEM solution (black dots) indicates that the exact scattering behavior of the spheroid ensemble is not correctly reproduced by the conventional T-matrix formalism. Such a mismatch is to be expected, since one particle's circumscribing sphere intersects the second particle. For the T-matrix simulation utilizing the PVWF coupling (orange line) we obtain a very good agreement with the FEM simulation. For the PWE, the integral over all in-plane wave numbers $\kappa$ (compare (\ref{equ:Wfinal})) has been considered up to the truncation value $\kappa_{\mathrm{trunc}}=3k$. For applicability reasons, the infinite integral has to be truncated at some finite value. As stated in our previous work \cite{Egel2016b}, one has to ensure, that for a fixed maximal multipole order $l_{\mathrm{max}}$ only values of the in-plane wave vector $\mathbf{k}_\parallel$ are considered, for which the angular power spectrum has converged against its true value. Very recently, a phenomenological formula for the estimation of $\kappa_{\mathrm{trunc}}$ has been proposed \cite{Egel2017}.

For a quantification of the accuracy of our simulation results, the relative deviation of both T-matrix formalisms from FEM-based solutions can be found in Fig. \ref{fig:2spheroidsplotgroup}(c). The relative deviation refers to the $L^2$-norm of the differential scattering cross sections and is shown for maximal multipole orders $l_{\mathrm{max}}=1$ up to $l_{\mathrm{max}}=20$, while the truncation of $\kappa$ is kept constant at $\kappa_{\mathrm{trunc}}=3k$. For low values of the maximal multipole order, a convergence of the angular power spectrum is not achieved for the fixed value of $\kappa_{\mathrm{trunc}} = 3k$. By increasing the maximal multipole order above $l_{\mathrm{max}}=7$, the relative deviation of the PVWF coupling formalism (orange dots) converges towards a minimal relative deviation of $1\%$. 

For the conventional superposition T-matrix formalism (blue circles) no convergence of the relative deviation can be obtained. Moreover, the relative deviation fluctuates around $10\%$, and strongly increases for large multipole orders $(l_{\mathrm{max}}\geq19)$. Such divergent behavior in the near field coupling has to be expected, as it reflects the divergence of the SWE in the near field with growing multipole order.

In a second example, we consider the two spheroids illustrated in Fig. \ref{fig:2spheroidsplotgroup}(a) to be made of silver with a refractive index of $n_p=0.13+2.918\mathrm{i}$ at $\lambda=500 \ \mathrm{nm}$ \cite{Palik}. Fig \ref{fig:2spheroidsplotgroup}(b) shows the DSCS for the silver spheroids at a maximal multipole order of $l_{\mathrm{max}}=15$ and $\kappa_{\mathrm{trunc}}=3k$. Again, the coupling via plane waves enables a good agreement with the FEM simulations, unlike the conventional superposition T-matrix formalism results, which strongly differ from the FEM reference. As shown in Fig. \ref{fig:2spheroidsplotgroup}(d), relative deviations, comparable to the $\mathrm{TiO_2}$-case, are obtained for the metallic nano-particles. Thus, above a multipole order of $l_{\mathrm{max}}=10$, the relative deviation starts converging towards $1.3 \ \%$ for the plane wave coupling, while it varies between $10 \ \%$ and $20 \ \%$ for the conventional formalism based on the spherical waves translation addition theorem.

\subsection{Cluster of spheroids}
\begin{figure*}
	\def\svgwidth{17.8cm}
	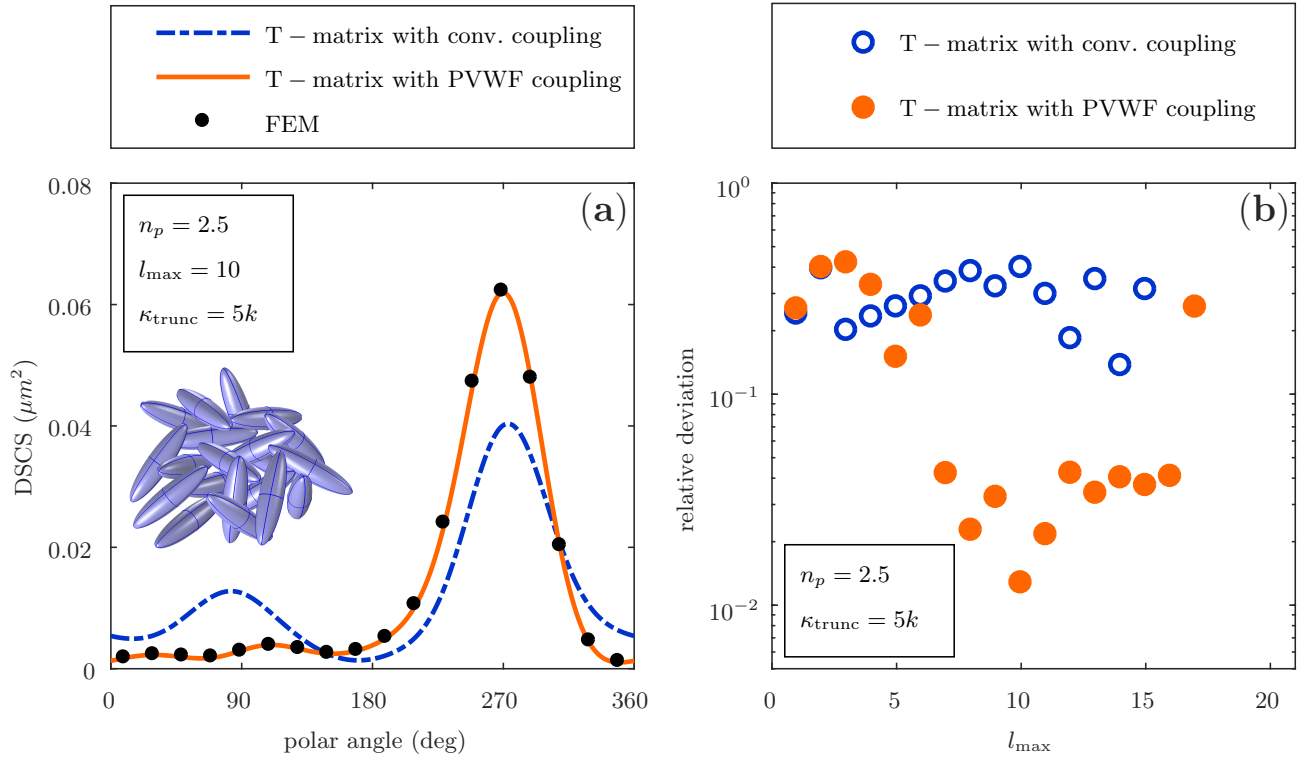
	\caption{\label{fig:spheroidcluster}Light scattering by 20 prolate spheroids $(a=b=30 \ \mathrm{nm}, c=120 \ \mathrm{nm})$. $\mathrm{TiO_2}$-particles $(n_p=2.5)$ are excited by a plane wave $(\lambda=500 \ \mathrm{nm})$, polarized in y-direction and propagating in negative z-direction. The ambient medium is air $(n_a=1)$. (a) Differential scattering cross-section of the spheroid cluster. For the SWE, multipole orders up to $l_{max}=10$ are considered. The PWE is truncated at $\kappa_{\mathrm{trunc}}=5k$. (b) shows the relative deviation of the DSCS for the conventional T-matrix formalism and the PVWF coupling procedure with respect to the FEM simulations. The maximal multipole order is varied from $l_{max}=1$ up to $l_{max}=20$, while the PWE truncation is kept constant at $\kappa_{\mathrm{trunc}}=5k$.}
\end{figure*}

In this example, we extend the validation of our approach to a more complex scattering system. The latter consists of a cluster made of 20 prolate $\mathrm{TiO_2}$-spheroids $(n_a=2.5)$ with semi-minor axes of $a=b=30 \ \mathrm{nm}$ and a semi-major axis of $c=120 \ \mathrm{nm}$. The cluster is formed by arbitrarily oriented particles (for visualization see Fig. \ref{fig:spheroidcluster}(a)). Such scattering clusters find applications in dye sensitized solar cells, where the $\mathrm{TiO_2}$-rods are exploited as a scattering layer for improving light-harvesting \cite{Fan2011,Shital2016}. Smaller, very dense clusters consisting of a few $\mathrm{TiO_2}$-particles can also be found in white paint, which can be used as light-trapping back-reflectors in photovoltaics \cite{Lipovsek2010}. 

The cluster considered herein functions as an extreme challenge for the plane wave coupling formalism. The prolate shape of the scattering particles and an aspect ratio of $4$ allows for very low distances between particle centers, in comparison to the particle diameters. In some cases, the high packing factor leads to a minimal distance between adjacent particles below $1 \ \mathrm{nm}$ and to overlapping of the circumscribing sphere of one particle with multiple neighboring particles. 

Figure \ref{fig:spheroidcluster}(a) compares the calculated DSCS of the spheroid cluster for the conventional superposition T-matrix scheme in conjunction with the translation addition theorem (blue-dotted line), the PVWF coupling formalism (orange line) and FEM simulations (black dots). The spherical wave expansion has been taken into account up to a multipole order of $l_{\mathrm{max}}=10$, while the plane wave expansion has been truncated at $\kappa_{\mathrm{trunc}}=5k$. A good agreement between the T-matrix simulations relying on the PVWF coupling formalism and the FEM can be observed, while the conventional T-matrix scheme's results do not match the FEM simulation. For $\kappa_{\mathrm{trunc}}=5k$ and large multipole orders $(l_{\mathrm{max}}>16)$, the relative deviation (see Fig. \ref{fig:spheroidcluster}(b)) between both T-matrix and the FEM simulations show divergent behavior. In this configuration, very low distances below $1 \ \mathrm{nm}$ lead to large values for the spherical Hankel function of first kind $h_l^{(1)}$ (see definition of the outgoing SVWFs (\ref{equ:svwf})). This can lead to an ill-conditioning of the linear system (\ref{equ:wmat}), when too large multipole orders $l_{\mathrm{max}}$ are considered. Such divergence has been reported for decreasing distances between a spheroid and an interface \cite{Doicu1999}. Doicu et al. state, that for each fixed distance a domain of the maximal multipole order $l_{\mathrm{max}}$ exists, for which small deviations in the computed scattering response are obtained.

Such a plateau can be observed for the relative deviation between the PVWF coupling formalism and FEM. For maximal multipole orders of $l_{\mathrm{max}}=7,...,16$, the relative deviation does not exceed a value of $4 \ \%$, with a minimal deviation of $1.3 \ \%$ at $l_{\mathrm{max}}=10$. In comparison, the conventional superposition T-matrix approach using the translation addition theorem for SVWFs shows a minimal deviation of $13.7 \ \%$ at $l_{\mathrm{max}}=14$ and typically exceeds $ 25\  \%$.

\section{Discussion and conclusions}
We have shown that the T-matrix approach can be suitable to evaluate light scattering by dense systems of highly non-spherical particles, even if the circumscribing spheres intersect adjacent particles. To account for multiple scattering of neighboring particles, the SVWF translation operator can be expressed in a plane wave expansion. In practice, one has to ensure that for a given maximal multipole order of the SWE, the in-plane wavenumber of the PWE is truncated in a regime, where the angular spectrum converges \cite{Egel2016b}. For low values of the maximal multipole order $l_{\mathrm{max}}$, the accuracy is limited by the multipole truncation error, whereas for large $l_{\mathrm{max}}$, the poor condition number of the linear system becomes prohibitive, compare \cite{Doicu1999,Egel2016b}.

In this contribution, we have applied the plane wave coupling formalism to the case of spheroids. In general, our approach works for any non-spherical particle with a convex surface shape (or arbitrary particles, as long as the convex hulls do not overlap).

Regarding the computation time, the presented formalism cannot compete with the conventional superposition T-matrix scheme using the SVWF translation addition theorem, but exceeds it by a factor of $10$ in terms of accuracy  for the shown examples. The additional effort of the plane wave coupling can be reduced to a minimum by utilizing the conventional scheme for coupling between particles that are not within a very low distance. This way, the additional effort brought by the plane wave coupling scheme scales only linear with the number of involved particles, and thereby becomes negligible with growing particle numbers.

We conclude that the range of applicability of the T-matrix approach is much larger than typically expected. It has the potential to solve light scattering problems in large disordered systems, where strictly numerical approaches like the FEM or the FDTD method struggle in terms of hardware requirements. 

\begin{acknowledgments}
DT and AE acknowledge support from the Karlsruhe School of Optics $\&$ Photonics (KSOP). GG gratefully acknowledges support from the Helmholtz Postdoc Program. This work was funded by the DFG through the priority program 1839.
\end{acknowledgments}

\appendix*
\section{Wave functions and transformations}
The plane vector wave functions are defined as:
\begin{equation}
\label{equ:planewave}
\mathbf{E}_j^\pm (\kappa, \alpha, \mathbf{r}) = \exp(\mathrm{i}\mathbf{k^\pm \cdot r}) \hat{\mathbf{e}}_j .
\end{equation}
Here, $(\kappa, \alpha, \pm k_z)$ define the cylindrical coordinates of the wave vector $\mathbf{k}^\pm$, with $k_z=\sqrt{k^2-\kappa^2}$ and the wave number $k=n_0\omega$. The plus sign corresponds to waves propagating in the positive z-direction, the minus sign refers to waves propagating in the negative z-direction. Index $j$ of $\mathbf{E}_j$ denotes the polarization (1 = TE and 2 = TM), realized by the unit vectors $\hat{\mathbf{e}}_1=\hat{\mathbf{e}}_\alpha$ and $\hat{\mathbf{e}}_2=\hat{\mathbf{e}}_\beta$, which belong to the azimuthal and polar angle of $\mathbf{k}^\pm$.

Besides plane wave functions, we make use of spherical vector wave functions, which read \cite{Doicu2006}:
\begin{align}
\label{equ:svwf}
&\mathbf{M}_{lm1}^{(\nu)}(\mathbf{r}) = \frac{1}{\sqrt{2l(l-1)}} \nabla \times \left(\mathbf{r}z_l^{(\nu)}(kr)P_l^{|m|}(\cos\theta)\mathrm{e}^{\mathrm{i}m\phi}\right) ,\\
&\mathbf{M}_{lm2}^{(\nu)}(\mathbf{r}) = \frac{1}{k} \nabla \times \mathbf{M}_{lm1}^{(\nu)}(\mathbf{r}),
\end{align}
where $(r,\theta, \phi)$ are the spherical coordinates of the position vector $\mathbf{r}$.
Index ($\nu$) indicates whether the wave function is of regular kind ($\nu=1$) or if it represents an outgoing wave ($\nu=3$). In case of regular spherical waves the radial wave function $z_l^{(\nu)}$ stands for the spherical Bessel function of order l, $z_l^{(1)}=j_l$. Outgoing spherical waves involve the spherical Hankel function of first kind, $z_l^{(3)}=h_l^{(1)}$. $P_l^{|m|}$ denotes the normalized associated Legendre functions. The spherical wave functions $\mathbf{M}_{lmp}^{(\nu)}$ are specified by the following indices: $l=1,2,...$ describes the angular index with respect to $\theta$, $m=-l,...,l$ the angular index with respect to $\phi$  and $p$ the spherical polarization ($1=\mathrm{TE}, 2=\mathrm{TM}$). For a more condensed notation the indices are subsumed into a multi index $(lmp) \rightarrow n$.

A translation of SVWFs can be accounted for by making use of the translation addition theorem \cite{Cruzan1962}:
\begin{equation}
\label{equ:translationadditiontheorem}
\mathbf{M}_n^{(3)}(\mathbf{r}+\mathbf{d}) = \displaystyle\sum_{n'} A_{nn'}(\mathbf{d}) \mathbf{M}_{n'}^{(1)}(\mathbf{r}) \ \ \mathrm{for}\ r < d .
\end{equation} 
The translation operator $A(\mathbf{d})$ can be obtained by making use of recurrence formulas given in \cite{Mackowski1991,Doicu2006}. Alternatively, one can use expressions involving the so called Wigner-3j symbols found e.g., in Ref(s). \cite{Cruzan1962,Stein1961,Mishchenko2002}. 

Transforming SVWFs from a laboratory coordinate system ($L$) to a rotated coordinate system ($R$) can be achieved, utilizing the rotation addition theorem for SVWFs \cite{Stein1961}:
\begin{equation}
\label{equ:rotationadditiontheorem}
\mathbf{M}_{lmp}^{(1,3)}(R) = \displaystyle\sum_{m'=-l}^{l} D_{mm'}^l(\alpha,\beta,\gamma) \mathbf{M}_{lm'p}^{(1,3)}(L) .
\end{equation}
Function $D$ refers to the so called Wigner D-functions. Recurrence formulations for $D$ can be found e.g., in Ref. \cite{Doicu2006} or Ref. \cite{Mishchenko2002}. The rotation between the two coordinate systems is defined by the Euler angles $(\alpha,\beta,\gamma)$ in $zy'z'$-convention. 

Spherical vector wave functions can be expanded in plane vector wave functions and vice versa \cite{Bostrom1991}:
\begin{align}
\label{equ:EtoM3}
\mathbf{M}_n^{(3)}(\mathbf{r}) = \ & \frac{1}{2\pi} \displaystyle\int_{\mathbb{R}^2}d^2\mathbf{k}_\parallel \frac{1}{k_zk} \displaystyle\sum_{j=1}^{2} B_{nj}\left(\frac{\pm k_z}{k}\right) \notag \\
& \times\mathbf{E}_j^\pm(\kappa,\alpha;\mathbf{r})\mathrm{e}^{\mathrm{i}m\alpha} \ \  \mathrm{for}\ z\gtrless0 . 
\end{align}
The integral is performed over the in-plane components of the wave vector $\mathbf{k}_\parallel$ with its polar coordinates $\kappa, \alpha$.

A plane wave representation in terms of regular spherical vector wave functions reads:
\begin{equation}
\label{equ:M1toE}
\mathbf{E}_j^\pm(\kappa,\alpha;\mathbf{r}) = 4\displaystyle\sum_{n} \mathrm{e}^{-\mathrm{i}m\alpha} B_{nj}^\dagger\left(\frac{\pm k_z}{k}\right) \mathbf{M}_n^{(1)}(\mathbf{r}) .
\end{equation}
The transformation operator $B$ is given by:
\begin{align}
\label{equ:Boperator}
B_{nj}(x) = &-\frac{1}{\mathrm{i}^{l+1}} \frac{1}{\sqrt{2l\left(l+1\right)}} \left(\mathrm{i}\delta_{j1}+\delta_{j2}\right) \notag \\
& \times \left(\delta_{pj}\tau_l^{|m|}(x) + (1-\delta_{pj})m\pi_l^{|m|}(x)\right) ,
\end{align}
where the spherical functions $\pi$ and $\tau$ are defined as:
\begin{align}
&\pi_l^m(\cos\theta) = \frac{P_l^m(\cos\theta)}{\sin\theta} , \notag \\
&\tau_l^m(\cos\theta) = \partial_\theta P_l^m(\cos\theta) . \notag
\end{align}
In the 'daggered' version of the transformation operator $B^\dagger$, all explicit $\mathrm{i}$ are set to $-\mathrm{i}$.

\bibliography{referencestheobald}

\begin{thebibliography}{34}%
\makeatletter
\providecommand \@ifxundefined [1]{%
 \@ifx{#1\undefined}
}%
\providecommand \@ifnum [1]{%
 \ifnum #1\expandafter \@firstoftwo
 \else \expandafter \@secondoftwo
 \fi
}%
\providecommand \@ifx [1]{%
 \ifx #1\expandafter \@firstoftwo
 \else \expandafter \@secondoftwo
 \fi
}%
\providecommand \natexlab [1]{#1}%
\providecommand \enquote  [1]{``#1''}%
\providecommand \bibnamefont  [1]{#1}%
\providecommand \bibfnamefont [1]{#1}%
\providecommand \citenamefont [1]{#1}%
\providecommand \href@noop [0]{\@secondoftwo}%
\providecommand \href [0]{\begingroup \@sanitize@url \@href}%
\providecommand \@href[1]{\@@startlink{#1}\@@href}%
\providecommand \@@href[1]{\endgroup#1\@@endlink}%
\providecommand \@sanitize@url [0]{\catcode `\\12\catcode `\$12\catcode
  `\&12\catcode `\#12\catcode `\^12\catcode `\_12\catcode `\%12\relax}%
\providecommand \@@startlink[1]{}%
\providecommand \@@endlink[0]{}%
\providecommand \url  [0]{\begingroup\@sanitize@url \@url }%
\providecommand \@url [1]{\endgroup\@href {#1}{\urlprefix }}%
\providecommand \urlprefix  [0]{URL }%
\providecommand \Eprint [0]{\href }%
\providecommand \doibase [0]{http://dx.doi.org/}%
\providecommand \selectlanguage [0]{\@gobble}%
\providecommand \bibinfo  [0]{\@secondoftwo}%
\providecommand \bibfield  [0]{\@secondoftwo}%
\providecommand \translation [1]{[#1]}%
\providecommand \BibitemOpen [0]{}%
\providecommand \bibitemStop [0]{}%
\providecommand \bibitemNoStop [0]{.\EOS\space}%
\providecommand \EOS [0]{\spacefactor3000\relax}%
\providecommand \BibitemShut  [1]{\csname bibitem#1\endcsname}%
\let\auto@bib@innerbib\@empty
\bibitem [{\citenamefont {Geier}\ and\ \citenamefont
  {Arienti}(2014)}]{geier2014}%
  \BibitemOpen
  \bibfield  {author} {\bibinfo {author} {\bibfnamefont {M.}~\bibnamefont
  {Geier}}\ and\ \bibinfo {author} {\bibfnamefont {M.}~\bibnamefont
  {Arienti}},\ }\href {\doibase 10.1016/j.jqsrt.2014.07.011} {\bibfield
  {journal} {\bibinfo  {journal} {J. Quant. Spectrosc. Radiat. Transf.}\
  }\textbf {\bibinfo {volume} {149}},\ \bibinfo {pages} {16} (\bibinfo {year}
  {2014})}\BibitemShut {NoStop}%
\bibitem [{\citenamefont {Tamanai}\ \emph {et~al.}(2006)\citenamefont
  {Tamanai}, \citenamefont {Mutschke}, \citenamefont {Blum},\ and\
  \citenamefont {Neuh{\"{a}}user}}]{Tamanai2006}%
  \BibitemOpen
  \bibfield  {author} {\bibinfo {author} {\bibfnamefont {A.}~\bibnamefont
  {Tamanai}}, \bibinfo {author} {\bibfnamefont {H.}~\bibnamefont {Mutschke}},
  \bibinfo {author} {\bibfnamefont {J.}~\bibnamefont {Blum}}, \ and\ \bibinfo
  {author} {\bibfnamefont {R.}~\bibnamefont {Neuh{\"{a}}user}},\ }\href@noop {}
  {\bibfield  {journal} {\bibinfo  {journal} {J. Quant. Spectrosc. Radiat.
  Transf.}\ }\textbf {\bibinfo {volume} {100}},\ \bibinfo {pages} {373}
  (\bibinfo {year} {2006})}\BibitemShut {NoStop}%
\bibitem [{\citenamefont {Wilts}\ \emph {et~al.}(2017)\citenamefont {Wilts},
  \citenamefont {Wijnen}, \citenamefont {Leertouwer}, \citenamefont {Steiner},\
  and\ \citenamefont {Stavenga}}]{Wilts2017}%
  \BibitemOpen
  \bibfield  {author} {\bibinfo {author} {\bibfnamefont {B.~D.}\ \bibnamefont
  {Wilts}}, \bibinfo {author} {\bibfnamefont {B.}~\bibnamefont {Wijnen}},
  \bibinfo {author} {\bibfnamefont {H.~L.}\ \bibnamefont {Leertouwer}},
  \bibinfo {author} {\bibfnamefont {U.}~\bibnamefont {Steiner}}, \ and\
  \bibinfo {author} {\bibfnamefont {D.~G.}\ \bibnamefont {Stavenga}},\
  }\href@noop {} {\bibfield  {journal} {\bibinfo  {journal} {Adv. Opt. Mater.}\
  }\textbf {\bibinfo {volume} {5}},\ \bibinfo {pages} {1600879} (\bibinfo
  {year} {2017})}\BibitemShut {NoStop}%
\bibitem [{\citenamefont {Dannhauser}\ \emph {et~al.}(2017)\citenamefont
  {Dannhauser}, \citenamefont {Rossi}, \citenamefont {Memmolo}, \citenamefont
  {Causa}, \citenamefont {Finizio}, \citenamefont {Ferraro},\ and\
  \citenamefont {Netti}}]{Dannhauser2017}%
  \BibitemOpen
  \bibfield  {author} {\bibinfo {author} {\bibfnamefont {D.}~\bibnamefont
  {Dannhauser}}, \bibinfo {author} {\bibfnamefont {D.}~\bibnamefont {Rossi}},
  \bibinfo {author} {\bibfnamefont {P.}~\bibnamefont {Memmolo}}, \bibinfo
  {author} {\bibfnamefont {F.}~\bibnamefont {Causa}}, \bibinfo {author}
  {\bibfnamefont {A.}~\bibnamefont {Finizio}}, \bibinfo {author} {\bibfnamefont
  {P.}~\bibnamefont {Ferraro}}, \ and\ \bibinfo {author} {\bibfnamefont
  {P.~A.}\ \bibnamefont {Netti}},\ }\href {\doibase 10.1002/jbio.201600070}
  {\bibfield  {journal} {\bibinfo  {journal} {J. Biophotonics}\ }\textbf
  {\bibinfo {volume} {10}},\ \bibinfo {pages} {683} (\bibinfo {year}
  {2017})}\BibitemShut {NoStop}%
\bibitem [{\citenamefont {Atwater}\ and\ \citenamefont
  {Polman}(2010)}]{Atwater2010}%
  \BibitemOpen
  \bibfield  {author} {\bibinfo {author} {\bibfnamefont {H.~A.}\ \bibnamefont
  {Atwater}}\ and\ \bibinfo {author} {\bibfnamefont {A.}~\bibnamefont
  {Polman}},\ }\href {\doibase 10.1038/nmat2866} {\bibfield  {journal}
  {\bibinfo  {journal} {Nat. Mater.}\ }\textbf {\bibinfo {volume} {9}},\
  \bibinfo {pages} {205} (\bibinfo {year} {2010})}\BibitemShut {NoStop}%
\bibitem [{\citenamefont {Gomard}\ \emph {et~al.}(2016)\citenamefont {Gomard},
  \citenamefont {Preinfalk}, \citenamefont {Egel},\ and\ \citenamefont
  {Lemmer}}]{Gomard2016}%
  \BibitemOpen
  \bibfield  {author} {\bibinfo {author} {\bibfnamefont {G.}~\bibnamefont
  {Gomard}}, \bibinfo {author} {\bibfnamefont {J.~B.}\ \bibnamefont
  {Preinfalk}}, \bibinfo {author} {\bibfnamefont {A.}~\bibnamefont {Egel}}, \
  and\ \bibinfo {author} {\bibfnamefont {U.}~\bibnamefont {Lemmer}},\
  }\href@noop {} {\bibfield  {journal} {\bibinfo  {journal} {J. Photonics
  Energy}\ }\textbf {\bibinfo {volume} {6}},\ \bibinfo {pages} {30901}
  (\bibinfo {year} {2016})}\BibitemShut {NoStop}%
\bibitem [{\citenamefont {Zhang}\ and\ \citenamefont {Cao}(2011)}]{Zhang2011}%
  \BibitemOpen
  \bibfield  {author} {\bibinfo {author} {\bibfnamefont {Q.}~\bibnamefont
  {Zhang}}\ and\ \bibinfo {author} {\bibfnamefont {G.}~\bibnamefont {Cao}},\
  }\href@noop {} {\bibfield  {journal} {\bibinfo  {journal} {Nano Today}\
  }\textbf {\bibinfo {volume} {6}},\ \bibinfo {pages} {91} (\bibinfo {year}
  {2011})}\BibitemShut {NoStop}%
\bibitem [{\citenamefont {Waterman}(1965)}]{Waterman1965}%
  \BibitemOpen
  \bibfield  {author} {\bibinfo {author} {\bibfnamefont {P.~C.}\ \bibnamefont
  {Waterman}},\ }\href {\doibase 10.1109/PROC.1965.4058} {\bibfield  {journal}
  {\bibinfo  {journal} {Proc. IEEE}\ }\textbf {\bibinfo {volume} {53}},\
  \bibinfo {pages} {805 } (\bibinfo {year} {1965})}\BibitemShut {NoStop}%
\bibitem [{\citenamefont {Doicu}\ \emph {et~al.}(2006)\citenamefont {Doicu},
  \citenamefont {Wriedt},\ and\ \citenamefont {Eremin}}]{Doicu2006}%
  \BibitemOpen
  \bibfield  {author} {\bibinfo {author} {\bibfnamefont {A.}~\bibnamefont
  {Doicu}}, \bibinfo {author} {\bibfnamefont {T.}~\bibnamefont {Wriedt}}, \
  and\ \bibinfo {author} {\bibfnamefont {Y.}~\bibnamefont {Eremin}},\ }\href
  {\doibase 10.1007/978-3-642-02646-1} {\emph {\bibinfo {title} {{Light
  Scattering by Systems of Particles}}}}\ (\bibinfo  {publisher} {Springer},\
  \bibinfo {address} {Berlin, Heidelberg},\ \bibinfo {year} {2006})\ p.\
  \bibinfo {pages} {322}\BibitemShut {NoStop}%
\bibitem [{\citenamefont {Mishchenko}\ \emph {et~al.}(2002)\citenamefont
  {Mishchenko}, \citenamefont {Travis},\ and\ \citenamefont
  {Lacis}}]{Mishchenko2002}%
  \BibitemOpen
  \bibfield  {author} {\bibinfo {author} {\bibfnamefont {M.~I.}\ \bibnamefont
  {Mishchenko}}, \bibinfo {author} {\bibfnamefont {L.~D.}\ \bibnamefont
  {Travis}}, \ and\ \bibinfo {author} {\bibfnamefont {A.~A.}\ \bibnamefont
  {Lacis}},\ }\href@noop {} {\emph {\bibinfo {title} {{Scattering, Absorption,
  and Emission of Light by Small Particles}}}}\ (\bibinfo  {publisher}
  {Cambridge University Press},\ \bibinfo {year} {2002})\ p.\ \bibinfo {pages}
  {462}\BibitemShut {NoStop}%
\bibitem [{\citenamefont {Cruzan}(1962)}]{Cruzan1962}%
  \BibitemOpen
  \bibfield  {author} {\bibinfo {author} {\bibfnamefont {O.~R.}\ \bibnamefont
  {Cruzan}},\ }\href@noop {} {\bibfield  {journal} {\bibinfo  {journal} {Q.
  Appl. Math.}\ }\textbf {\bibinfo {volume} {20}},\ \bibinfo {pages} {33}
  (\bibinfo {year} {1962})}\BibitemShut {NoStop}%
\bibitem [{\citenamefont {Peterson}\ and\ \citenamefont
  {Str{\"{o}}m}(1973)}]{Peterson1973}%
  \BibitemOpen
  \bibfield  {author} {\bibinfo {author} {\bibfnamefont {B.}~\bibnamefont
  {Peterson}}\ and\ \bibinfo {author} {\bibfnamefont {S.}~\bibnamefont
  {Str{\"{o}}m}},\ }\href {\doibase 10.1103/PhysRevD.8.3661} {\bibfield
  {journal} {\bibinfo  {journal} {Phys. Rev. D}\ }\textbf {\bibinfo {volume}
  {8}},\ \bibinfo {pages} {3661} (\bibinfo {year} {1973})}\BibitemShut
  {NoStop}%
\bibitem [{\citenamefont {Varadan}\ and\ \citenamefont
  {Varadan}(1980)}]{Varadan1980}%
  \BibitemOpen
  \bibfield  {author} {\bibinfo {author} {\bibfnamefont {V.~V.}\ \bibnamefont
  {Varadan}}\ and\ \bibinfo {author} {\bibfnamefont {V.~K.}\ \bibnamefont
  {Varadan}},\ }\href@noop {} {\bibfield  {journal} {\bibinfo  {journal} {Phys.
  Rev. D}\ }\textbf {\bibinfo {volume} {21}},\ \bibinfo {pages} {388} (\bibinfo
  {year} {1980})}\BibitemShut {NoStop}%
\bibitem [{\citenamefont {Mishchenko}\ \emph {et~al.}(1996)\citenamefont
  {Mishchenko}, \citenamefont {Travis},\ and\ \citenamefont
  {Mackowski}}]{Mishchenko1996}%
  \BibitemOpen
  \bibfield  {author} {\bibinfo {author} {\bibfnamefont {M.~I.}\ \bibnamefont
  {Mishchenko}}, \bibinfo {author} {\bibfnamefont {L.~D.}\ \bibnamefont
  {Travis}}, \ and\ \bibinfo {author} {\bibfnamefont {D.~W.}\ \bibnamefont
  {Mackowski}},\ }\href@noop {} {\bibfield  {journal} {\bibinfo  {journal} {J.
  Quant. Spectrosc. Radial. Transf.}\ }\textbf {\bibinfo {volume} {55}},\
  \bibinfo {pages} {535} (\bibinfo {year} {1996})}\BibitemShut {NoStop}%
\bibitem [{\citenamefont {Wriedt}\ \emph {et~al.}(2008)\citenamefont {Wriedt},
  \citenamefont {Schuh},\ and\ \citenamefont {Doicu}}]{Wriedt2008a}%
  \BibitemOpen
  \bibfield  {author} {\bibinfo {author} {\bibfnamefont {T.}~\bibnamefont
  {Wriedt}}, \bibinfo {author} {\bibfnamefont {R.}~\bibnamefont {Schuh}}, \
  and\ \bibinfo {author} {\bibfnamefont {A.}~\bibnamefont {Doicu}},\
  }\href@noop {} {\bibfield  {journal} {\bibinfo  {journal} {Part. Part. Syst.
  Charact.}\ }\textbf {\bibinfo {volume} {25}},\ \bibinfo {pages} {74}
  (\bibinfo {year} {2008})}\BibitemShut {NoStop}%
\bibitem [{\citenamefont {Egel}\ \emph {et~al.}(2016)\citenamefont {Egel},
  \citenamefont {Theobald}, \citenamefont {Donie}, \citenamefont {Lemmer},\
  and\ \citenamefont {Gomard}}]{Egel2016b}%
  \BibitemOpen
  \bibfield  {author} {\bibinfo {author} {\bibfnamefont {A.}~\bibnamefont
  {Egel}}, \bibinfo {author} {\bibfnamefont {D.}~\bibnamefont {Theobald}},
  \bibinfo {author} {\bibfnamefont {Y.}~\bibnamefont {Donie}}, \bibinfo
  {author} {\bibfnamefont {U.}~\bibnamefont {Lemmer}}, \ and\ \bibinfo {author}
  {\bibfnamefont {G.}~\bibnamefont {Gomard}},\ }\href@noop {} {\bibfield
  {journal} {\bibinfo  {journal} {Opt. Express}\ }\textbf {\bibinfo {volume}
  {24}},\ \bibinfo {pages} {25154} (\bibinfo {year} {2016})}\BibitemShut
  {NoStop}%
\bibitem [{\citenamefont {Shital}\ and\ \citenamefont
  {Dutta}(2016)}]{Shital2016}%
  \BibitemOpen
  \bibfield  {author} {\bibinfo {author} {\bibfnamefont {S.}~\bibnamefont
  {Shital}}\ and\ \bibinfo {author} {\bibfnamefont {V.}~\bibnamefont {Dutta}},\
  }\href@noop {} {\bibfield  {journal} {\bibinfo  {journal} {J. Photonics
  Energy}\ }\textbf {\bibinfo {volume} {6}},\ \bibinfo {pages} {25503}
  (\bibinfo {year} {2016})}\BibitemShut {NoStop}%
\bibitem [{\citenamefont {Wriedt}\ and\ \citenamefont
  {Hellmers}(2008)}]{Wriedt2008}%
  \BibitemOpen
  \bibfield  {author} {\bibinfo {author} {\bibfnamefont {T.}~\bibnamefont
  {Wriedt}}\ and\ \bibinfo {author} {\bibfnamefont {J.}~\bibnamefont
  {Hellmers}},\ }\href@noop {} {\bibfield  {journal} {\bibinfo  {journal} {J.
  Quant. Spectrosc. Radiat. Transf.}\ }\textbf {\bibinfo {volume} {109}},\
  \bibinfo {pages} {1536} (\bibinfo {year} {2008})}\BibitemShut {NoStop}%
\bibitem [{\citenamefont {Liu}\ \emph {et~al.}(2008)\citenamefont {Liu},
  \citenamefont {Mishchenko},\ and\ \citenamefont {{Patrick
  Arnott}}}]{Liu2008}%
  \BibitemOpen
  \bibfield  {author} {\bibinfo {author} {\bibfnamefont {L.}~\bibnamefont
  {Liu}}, \bibinfo {author} {\bibfnamefont {M.~I.}\ \bibnamefont {Mishchenko}},
  \ and\ \bibinfo {author} {\bibfnamefont {W.}~\bibnamefont {{Patrick
  Arnott}}},\ }\href@noop {} {\bibfield  {journal} {\bibinfo  {journal} {J.
  Quant. Spectrosc. Radiat. Transf.}\ }\textbf {\bibinfo {volume} {109}},\
  \bibinfo {pages} {2656} (\bibinfo {year} {2008})}\BibitemShut {NoStop}%
\bibitem [{\citenamefont {Mackowski}\ and\ \citenamefont
  {Mishchenko}(2011)}]{Mackowski2011}%
  \BibitemOpen
  \bibfield  {author} {\bibinfo {author} {\bibfnamefont {D.~W.}\ \bibnamefont
  {Mackowski}}\ and\ \bibinfo {author} {\bibfnamefont {M.~I.}\ \bibnamefont
  {Mishchenko}},\ }\href@noop {} {\bibfield  {journal} {\bibinfo  {journal}
  {Phys. Rev. A}\ }\textbf {\bibinfo {volume} {83}},\ \bibinfo {pages} {013804}
  (\bibinfo {year} {2011})}\BibitemShut {NoStop}%
\bibitem [{\citenamefont {Egel}\ and\ \citenamefont {Lemmer}(2014)}]{Egel2014}%
  \BibitemOpen
  \bibfield  {author} {\bibinfo {author} {\bibfnamefont {A.}~\bibnamefont
  {Egel}}\ and\ \bibinfo {author} {\bibfnamefont {U.}~\bibnamefont {Lemmer}},\
  }\href {\doibase 10.1016/j.jqsrt.2014.06.022} {\bibfield  {journal} {\bibinfo
   {journal} {J. Quant. Spectrosc. Radiat. Transf.}\ }\textbf {\bibinfo
  {volume} {148}},\ \bibinfo {pages} {165} (\bibinfo {year}
  {2014})}\BibitemShut {NoStop}%
\bibitem [{\citenamefont {Fruhnert}\ \emph {et~al.}(2016)\citenamefont
  {Fruhnert}, \citenamefont {Monti}, \citenamefont {Fernandez-Corbaton},
  \citenamefont {Al{\`{u}}}, \citenamefont {Toscano}, \citenamefont {Bilotti},\
  and\ \citenamefont {Rockstuhl}}]{Fruhnert2016}%
  \BibitemOpen
  \bibfield  {author} {\bibinfo {author} {\bibfnamefont {M.}~\bibnamefont
  {Fruhnert}}, \bibinfo {author} {\bibfnamefont {A.}~\bibnamefont {Monti}},
  \bibinfo {author} {\bibfnamefont {I.}~\bibnamefont {Fernandez-Corbaton}},
  \bibinfo {author} {\bibfnamefont {A.}~\bibnamefont {Al{\`{u}}}}, \bibinfo
  {author} {\bibfnamefont {A.}~\bibnamefont {Toscano}}, \bibinfo {author}
  {\bibfnamefont {F.}~\bibnamefont {Bilotti}}, \ and\ \bibinfo {author}
  {\bibfnamefont {C.}~\bibnamefont {Rockstuhl}},\ }\href@noop {} {\bibfield
  {journal} {\bibinfo  {journal} {Phys. Rev. B}\ }\textbf {\bibinfo {volume}
  {93}},\ \bibinfo {pages} {245127} (\bibinfo {year} {2016})}\BibitemShut
  {NoStop}%
\bibitem [{\citenamefont {Markkanen}\ and\ \citenamefont
  {Yuffa}(2017)}]{Markkanen2017}%
  \BibitemOpen
  \bibfield  {author} {\bibinfo {author} {\bibfnamefont {J.}~\bibnamefont
  {Markkanen}}\ and\ \bibinfo {author} {\bibfnamefont {A.~J.}\ \bibnamefont
  {Yuffa}},\ }\href@noop {} {\bibfield  {journal} {\bibinfo  {journal} {J.
  Quant. Spectrosc. Radiat. Transf.}\ }\textbf {\bibinfo {volume} {189}},\
  \bibinfo {pages} {181} (\bibinfo {year} {2017})}\BibitemShut {NoStop}%
\bibitem [{\citenamefont {Bates}(1975)}]{Bates1975}%
  \BibitemOpen
  \bibfield  {author} {\bibinfo {author} {\bibfnamefont {R.~H.~T.}\
  \bibnamefont {Bates}},\ }\href {\doibase 10.1109/TMTT.1975.1128639}
  {\bibfield  {journal} {\bibinfo  {journal} {IEEE Trans. Microw. Theory
  Tech.}\ }\textbf {\bibinfo {volume} {23}},\ \bibinfo {pages} {605} (\bibinfo
  {year} {1975})}\BibitemShut {NoStop}%
\bibitem [{\citenamefont {Doicu}\ \emph {et~al.}(1999)\citenamefont {Doicu},
  \citenamefont {Eremin},\ and\ \citenamefont {Wriedt}}]{Doicu1999}%
  \BibitemOpen
  \bibfield  {author} {\bibinfo {author} {\bibfnamefont {A.}~\bibnamefont
  {Doicu}}, \bibinfo {author} {\bibfnamefont {Y.~A.}\ \bibnamefont {Eremin}}, \
  and\ \bibinfo {author} {\bibfnamefont {T.}~\bibnamefont {Wriedt}},\
  }\href@noop {} {\bibfield  {journal} {\bibinfo  {journal} {Opt. Commun.}\
  }\textbf {\bibinfo {volume} {159}},\ \bibinfo {pages} {266} (\bibinfo {year}
  {1999})}\BibitemShut {NoStop}%
\bibitem [{\citenamefont {Bostr{\"{o}}m}\ \emph {et~al.}(1991)\citenamefont
  {Bostr{\"{o}}m}, \citenamefont {Kristensson},\ and\ \citenamefont
  {Str{\"{o}}m}}]{Bostrom1991}%
  \BibitemOpen
  \bibfield  {author} {\bibinfo {author} {\bibfnamefont {A.}~\bibnamefont
  {Bostr{\"{o}}m}}, \bibinfo {author} {\bibfnamefont {G.}~\bibnamefont
  {Kristensson}}, \ and\ \bibinfo {author} {\bibfnamefont {S.}~\bibnamefont
  {Str{\"{o}}m}},\ }in\ \href@noop {} {\emph {\bibinfo {booktitle} {F.
  Represent. Introd. to Scatt.}}}\ (\bibinfo  {publisher} {Elsevier Science
  Publishers B.V.},\ \bibinfo {year} {1991})\ pp.\ \bibinfo {pages}
  {165--210}\BibitemShut {NoStop}%
\bibitem [{\citenamefont {Stratton}(1941)}]{Adams}%
  \BibitemOpen
  \bibfield  {author} {\bibinfo {author} {\bibfnamefont {J.~A.}\ \bibnamefont
  {Stratton}},\ }\href@noop {} {\emph {\bibinfo {title} {{Electromagnetic
  Theory}}}}\ (\bibinfo  {publisher} {McGraw-Hill Book Company},\ \bibinfo
  {address} {New York and London},\ \bibinfo {year} {1941})\ p.\ \bibinfo
  {pages} {648}\BibitemShut {NoStop}%
\bibitem [{\citenamefont {Jalava}\ \emph {et~al.}(2015)\citenamefont {Jalava},
  \citenamefont {Taavitsainen}, \citenamefont {Lamminm{\"{a}}ki}, \citenamefont
  {Lindholm}, \citenamefont {Auvinen}, \citenamefont {Alatalo}, \citenamefont
  {Vartiainen},\ and\ \citenamefont {Haario}}]{Jalava2015}%
  \BibitemOpen
  \bibfield  {author} {\bibinfo {author} {\bibfnamefont {J.~P.}\ \bibnamefont
  {Jalava}}, \bibinfo {author} {\bibfnamefont {V.~M.}\ \bibnamefont
  {Taavitsainen}}, \bibinfo {author} {\bibfnamefont {R.~J.}\ \bibnamefont
  {Lamminm{\"{a}}ki}}, \bibinfo {author} {\bibfnamefont {M.}~\bibnamefont
  {Lindholm}}, \bibinfo {author} {\bibfnamefont {S.}~\bibnamefont {Auvinen}},
  \bibinfo {author} {\bibfnamefont {M.}~\bibnamefont {Alatalo}}, \bibinfo
  {author} {\bibfnamefont {E.}~\bibnamefont {Vartiainen}}, \ and\ \bibinfo
  {author} {\bibfnamefont {H.}~\bibnamefont {Haario}},\ }\href {\doibase
  10.1016/j.jqsrt.2015.08.007} {\bibfield  {journal} {\bibinfo  {journal} {J.
  Quant. Spectrosc. Radiat. Transf.}\ }\textbf {\bibinfo {volume} {167}},\
  \bibinfo {pages} {105} (\bibinfo {year} {2015})}\BibitemShut {NoStop}%
\bibitem [{\citenamefont {Egel}\ \emph {et~al.}()\citenamefont {Egel},
  \citenamefont {Eremin}, \citenamefont {Wriedt}, \citenamefont {Theobald},
  \citenamefont {Lemmer},\ and\ \citenamefont {Gomard}}]{Egel2017}%
  \BibitemOpen
  \bibfield  {author} {\bibinfo {author} {\bibfnamefont {A.}~\bibnamefont
  {Egel}}, \bibinfo {author} {\bibfnamefont {Y.}~\bibnamefont {Eremin}},
  \bibinfo {author} {\bibfnamefont {T.}~\bibnamefont {Wriedt}}, \bibinfo
  {author} {\bibfnamefont {D.}~\bibnamefont {Theobald}}, \bibinfo {author}
  {\bibfnamefont {U.}~\bibnamefont {Lemmer}}, \ and\ \bibinfo {author}
  {\bibfnamefont {G.}~\bibnamefont {Gomard}},\ }\href {\doibase
  10.1016/j.jqsrt.2017.08.016} {\bibfield  {journal} {\bibinfo  {journal} {J.
  Quant. Spectrosc. Radiat. Transf.}\ }10.1016/j.jqsrt.2017.08.016},\ \bibinfo
  {note} {(in press)}\BibitemShut {NoStop}%
\bibitem [{\citenamefont {Palik}(1997)}]{Palik}%
  \BibitemOpen
  \bibfield  {author} {\bibinfo {author} {\bibfnamefont {E.~D.}\ \bibnamefont
  {Palik}},\ }\href@noop {} {\emph {\bibinfo {title} {{Handbook of Optical
  Constants of Solids}}}}\ (\bibinfo  {publisher} {Academic Press},\ \bibinfo
  {address} {San Diego},\ \bibinfo {year} {1997})\ p.\ \bibinfo {pages}
  {325}\BibitemShut {NoStop}%
\bibitem [{\citenamefont {Fan}\ \emph {et~al.}(2011)\citenamefont {Fan},
  \citenamefont {Zhang}, \citenamefont {Peng}, \citenamefont {Chen},\ and\
  \citenamefont {Yang}}]{Fan2011}%
  \BibitemOpen
  \bibfield  {author} {\bibinfo {author} {\bibfnamefont {K.}~\bibnamefont
  {Fan}}, \bibinfo {author} {\bibfnamefont {W.}~\bibnamefont {Zhang}}, \bibinfo
  {author} {\bibfnamefont {T.}~\bibnamefont {Peng}}, \bibinfo {author}
  {\bibfnamefont {J.}~\bibnamefont {Chen}}, \ and\ \bibinfo {author}
  {\bibfnamefont {F.}~\bibnamefont {Yang}},\ }\href {\doibase
  10.1021/jp204725f} {\bibfield  {journal} {\bibinfo  {journal} {J. Phys. Chem.
  C}\ }\textbf {\bibinfo {volume} {115}},\ \bibinfo {pages} {17213} (\bibinfo
  {year} {2011})}\BibitemShut {NoStop}%
\bibitem [{\citenamefont {Lipov{\v{s}}ek}\ \emph {et~al.}(2010)\citenamefont
  {Lipov{\v{s}}ek}, \citenamefont {Kr{\v{c}}}, \citenamefont {Isabella},
  \citenamefont {Zeman},\ and\ \citenamefont {Topi}}]{Lipovsek2010}%
  \BibitemOpen
  \bibfield  {author} {\bibinfo {author} {\bibfnamefont {B.}~\bibnamefont
  {Lipov{\v{s}}ek}}, \bibinfo {author} {\bibfnamefont {J.}~\bibnamefont
  {Kr{\v{c}}}}, \bibinfo {author} {\bibfnamefont {O.}~\bibnamefont {Isabella}},
  \bibinfo {author} {\bibfnamefont {M.}~\bibnamefont {Zeman}}, \ and\ \bibinfo
  {author} {\bibfnamefont {M.}~\bibnamefont {Topi}},\ }\href@noop {} {\bibfield
   {journal} {\bibinfo  {journal} {J. Appl. Phys.}\ }\textbf {\bibinfo {volume}
  {108}},\ \bibinfo {pages} {103115} (\bibinfo {year} {2010})}\BibitemShut
  {NoStop}%
\bibitem [{\citenamefont {Mackowski}(1991)}]{Mackowski1991}%
  \BibitemOpen
  \bibfield  {author} {\bibinfo {author} {\bibfnamefont {D.~A.}\ \bibnamefont
  {Mackowski}},\ }\href@noop {} {\bibfield  {journal} {\bibinfo  {journal}
  {Proc. Math. Phys. Sci.}\ }\textbf {\bibinfo {volume} {433}},\ \bibinfo
  {pages} {599} (\bibinfo {year} {1991})}\BibitemShut {NoStop}%
\bibitem [{\citenamefont {Stein}(1961)}]{Stein1961}%
  \BibitemOpen
  \bibfield  {author} {\bibinfo {author} {\bibfnamefont {S.}~\bibnamefont
  {Stein}},\ }\href {\doibase 10.2144/000113897} {\bibfield  {journal}
  {\bibinfo  {journal} {Q. Appl. Math.}\ }\textbf {\bibinfo {volume} {19}},\
  \bibinfo {pages} {15} (\bibinfo {year} {1961})}\BibitemShut {NoStop}%
\end{thebibliography}%

\end{document}